# Frustration Driven Stripe Domain Formation in Co/Pt Multilayer Films

J. E. Davies,[1,2] O. Hellwig[3], E. E. Fullerton[4], M. Winklhofer[2,5], R. D. Shull[1], and Kai Liu[2,*]

[1]*Metallurgy Division, National Institute of Standards and Technology, Gaithersburg, MD 20899*

[2]*Physics Department, University of California, Davis, CA 95616*

[3]*Hitachi Global Storage Technologies, San Jose, CA*

[4]*Center for Magnetic Recording Research, University of California- San Diego, La Jolla, CA 92093-0401*

[5]*Department of Geosciences, University of Munich, D-80333, München, Germany*

## Abstract

We report microscopic mechanisms for an unusual magnetization reversal behavior in Co/Pt multilayers where some of the first-order reversal curves protrude outside of the major loop. Transmission x-ray microscopy reveals a fragmented stripe domain topography when the magnetic field is reversed prior to saturation, in contrast to an interconnected pattern when reversing from a saturated state. The different domain nucleation and propagation behaviors are due to unannihilated domains from the prior field sweep. These residual domains contribute to random dipole fields that impede the subsequent domain growth and prevent domains from growing as closely together as for the interconnected pattern.

**PACS: 75.60.-d, 75.60.Ch, 75.70.Kw, 75.70.Cn**

Magnetic thin films with strong perpendicular anisotropy have been extensively studied for their fascinating magnetization reversal processes and important applications in perpendicular magnetic recording.[1-7] These films typically undergo macroscopic magnetization reversal by nucleation and growth of microscopic domains. More generally, similar domain formation is observed in a wide range of other materials such as polymers, superconductors, liquid crystals and ferrofluids,[8, 9] which have drastically different interactions that govern their behaviors. The domain topographies resulting from these interactions and from external effects such as field cycling, temperature and strain may possess similarities within this diverse class of systems, and can certainly impact their macroscopic behavior.

Recent studies on perpendicular-anisotropy films have highlighted the role of the film microstructures on the macroscopic magnetization reversal.[5, 10-15] The presence of structural defects lead to numerous local minima in the energy landscape of the system. Under magnetic field cycling, these minima and the whole energy landscape evolve, which in turn determine the path the system takes to reconfigure. While it is common that the magnetic configurations depend on whether the system is cycled along the major or a minor loop,[11] in most cases the minor loops themselves lie inside the major loop. In this Letter we demonstrate a frustrated domain growth mechanism that leads to an unusual behavior in [Co(4Å)/Pt(7Å)]$_X$ multilayers where part of the minor loops actually protrude outside of the major loop.

The samples under investigation are Pt(200 Å)/[Co(4Å)/Pt(7Å)]$_X$ multilayer thin films with varying $X$. The samples were grown by magnetron sputtering at 0.4 Pa (3 mtorr) Ar pressure and ambient temperature onto Si$_3$N$_x$ coated Si substrates as well as Si$_3$N$_x$ membranes for transmission x-ray microscopy (TXRM) measurements. After deposition the films were capped with a 20 Å Pt layer. More details of the sample preparation can be found elsewhere.[5, 10, 14]

Magnetic properties were studied using an alternating gradient magnetometer (AGM) at room temperature with the applied field perpendicular to the film. For detailed study of magnetization reversal behaviors we have employed a FORC technique which quantifies the irreversible components of the magnetization with a FORC distribution: $\rho \equiv -\partial^2 M(H, H_R)/2\partial H \partial H_R$, as described earlier.[10, 16, 17] All measurements have been done with the same field sweep rate and dwell time after field-setting. TXRM observation of the magnetization reversal process was obtained using the XM-1 zone-plate imaging microscope on beamline 6.1.2 at the Lawrence Berkeley National Laboratory - Advanced Light Source (ALS). The x-ray energy was tuned to the Co $L_3$ absorption edge (778 eV). Additionally, Magneto-optical Kerr microscopy studies were carried out at the National Institute of Standards and Technology.

A family of FORC's and the corresponding FORC distribution for the $X = 20$ sample are shown in Figs. 1(a) and 1(b) respectively. The value of $\rho(\mu_o H_R, \mu_o H)$ is represented by different contour shadings and reveals three distinct reversal stages:[10] a horizontal ridge for -0.02 T < $\mu_o H_R$ < 0 corresponds to an initial rapid and irreversible domain propagation from already nucleated sites, which corresponds to a precipitous drop in the magnetization; a featureless plateau with $\rho \approx 0$ for -0.10 T< $\mu_o H_R$ < -0.02 T due to a second stage of mostly reversible labyrinth domain expansion/contraction without significant change of the domain morphology; and finally a vertical negative/positive pair of peaks for -0.25 T< $\mu_o H_R$ < -0.10 T due to the irreversible domain annihilation process. While details of these features vary with $X$ (e.g., the nucleation and saturation fields, and the size of the $\rho \approx 0$ plateau), qualitatively similar FORC distributions have been observed in samples with $X$ ranging from 10 to 150.

Close inspection of Fig. 1(a) reveal that while most of the FORC's reside within the major loop, some actually *protrude outside of the ascending-field branch of the major loop*. This unusual behavior is observed in samples with $X$ ranging from 10 to 50 and is more explicitly shown in Figs. 2(a) and 2(c) for samples with $X = 20$ and 50, respectively. Highlighted in these figures are three key FORC's corresponding to (*i*) where the protrusion begins, (*ii*) where the protrusion is a maximum and (*iii*) reversal from negative saturation. To better illustrate the reversal stages at which FORC's (*i*) – (*iii*) occur, the corresponding $\mu_o H_R$ values are marked as points (*i*) – (*iii*) on the FORC-derived switching field distribution (FORC-SFD) in Figs. 2(b) and 2(d). The FORC-SFD was obtained by projecting $\rho$ onto the $\mu_o H_R$ axis and yields the amount of irreversible switching that has occurred at a given $\mu_o H_R$.[18]

For $X = 20$ the protrusion begins with the FORC reversing at $\mu_o H_R = -62$ mT ($M/M_s = -0.32$) [FORC (*i*) in Fig. 2(a)]. This value of $\mu_o H_R$ is within the second stage of reversal, dominated by mostly reversible domain expansion [point (*i*) in Fig. 2(b)]. At more negative $\mu_o H_R$ the FORC's continue to extrude further outside the major loop until $\mu_o H_R = -135$ mT ($M/M_s = -0.85$) [FORC (*ii*) in Fig. 2(a)], where the reversal is dominated by domain annihilation [point (ii) in Fig. 2(b)]. Further reducing $\mu_o H_R$ towards negative saturation, the FORC's begin to conform onto the ascending-field branch of the major loop [FORC (*iii*) in Fig. 2(a)] and the FORC-SFD becomes zero [point (*iii*) in Fig. 2(b)]. A similar trend is observed for $X = 50$ where the protrusion also begins during the reversible domain expansion/contraction stage, and the maximum protrusion occurs during the annihilation stage [FORC/point (*ii*) in Figs. 2(c) and 2(d)] with $M/M_s \sim -0.85$. One interesting difference to note is for $X = 50$ the FORC's cross-over the major loop at negative fields compared to positive fields for $X = 20$. This difference is manifested in a monotonic DC-demagnetization remanence curve for $X = 20$ and a non-monotonic one for $X = 50$ [dashed line in Figs. 2(b) and 2(d) respectively].

Although the appearance of minor loops or FORC's crossing-over the major loop has been observed before,[19, 20] the underlying mechanism is either unknown[19] or not applicable to the present system.[20] Thus we have employed TXRM to correlate the observed macroscopic behavior from FORC with the microscopic processes. Figs. 3(a) and 3(b) show two TXRM micrographs of the magnetic domain structure for $X = 50$ over the same sample area, taken at applied fields of: (a) 90 mT and (b) 96 mT after exposing the sample to a maximum field ($\mu_o H_{max}$) of 350 mT and 306 mT, respectively.[21] The dark features correspond to reversed domains. For $\mu_o H_{max} = 350$ mT, the magnetization reversal starts from a completely saturated state; as the field is reduced reversal domains nucleate from only a few sites and propagate unabated through the sample, forming large and interconnected labyrinth domain networks. On the other hand the domain pattern for $\mu_o H_{max} = 306$ mT is largely disconnected, consisting of about 45 separate domain networks over the same sample area [Fig. 3(b)]. The key difference is that the $\mu_o H_{max} = 306$ mT is not enough to fully saturate the sample. This leaves behind many residual domains that become the nucleation sites for reverse domains during the subsequent field sweep.[10] Along a given FORC, the size of each interconnected labyrinth domain grown from a residual domain is limited by the separation between the adjacent residual domains due to dipole fields; after the reversed domains have propagated throughout the sample (e.g., near zero applied field for this $X=50$ sample), the up and down domains balance out each other except for frustrated moments at the end points of the labyrinth domains. The result is an increased net magnetization. As $\mu_o H_R$ approaches negative saturation (a later FORC), the number of residual domains decrease[10] and the size of the interconnected labyrinth networks increases; consequently the total frustrated moments reduce and a smaller net magnetization is observed. The overall effect is the protrusion of the FORC's outside of the major loop.

The differences in the domain pattern for $\mu_oH_{max}$ = 306 mT compared to $\mu_oH_{max}$ = 350 mT also extend to a difference in the number of topological defects in the domain structure.[9] Here we define topological defects as "end points" where domain growth has terminated and "branch points" where domains branch into at least three directions. The domain pattern for $\mu_oH_{max}$ = 306 mT has an 18 % increase in end points and 59 % decrease in branch points compared to $\mu_oH_{max}$ = 350 mT. This difference is another indication that there is restriction of the domain growth due to frustration. In particular, the substantial decrease in branch points shows that domains are unable to branch out to maximize the areal coverage and are instead restricted to single linear patterns. Additionally, approximately 40 topological defects within the viewing area (corresponding to $1.8 \times 10^8$ defects / cm$^2$) remain at the same locations for both reversal fields, potentially caused by structural defects (i.e. pinning sites) in the film.

It is worth noting that the history-dependent domain nucleation process does not always lead to protruding FORC's. Figs. 3(c) and 3(d) show Kerr microscopy studies of a sample with $X$ = 2, which does not exhibit the aforementioned crossover behavior in the FORC's. The sample is first saturated in a field of $\mu_oH_{max}$ = 15 mT. When a reversal field of -3.5 mT is applied, a single reverse domain nucleates and propagates through the film with increasing fields [Fig. 3(c)]. When negative saturation is reached at -15 mT and the field is cycled through a new $\mu_oH_{max}$ = +3.5 mT and brought back to -3.5 mT, multiple additional domains have nucleated [Fig. 3(d)]. Note that the domains during reversal are much larger than those in the $X$ = 50 sample.

The different reversal characteristics can be understood by considering the corresponding energies involved. As shown earlier,[14] due to the competition between magnetostatic and domain wall energies, the domain size in these [Co(4Å)/Pt(7Å)]$_X$ films depends on $X$. For samples with $X$ < 10 the energy gain forming a domain state is so small and the sample prefers to be uniformly magnetized and reversed by large micron-sized domains. This results in square major loops with abrupt switching once nucleation has begun and no cross-over behavior is observed. For $X > 10$ there is significant energy gain forming the stripe phase. However, during the reversal process, the disconnected and irregular labyrinth domain patterns cause random dipole fields which *impede* the domain growth from re-nucleated domains, preventing the domains from dense packing and growing together as closely as the interconnected domain structures. As a result, the different amount of *frustrated* moments gives rise to the protruded FORC's. Micromagnetic simulations[22] have shown that along protruding portions of the FORC's the total energy is greater than that along the ascending-field branch of the major loop.[23] This metastability of the microscopic magnetic configurations is to be expected in a frustrated system.

In summary we have demonstrated a microscopic mechanism for protruded FORC's observed in $(Co/Pt)_X$ multilayers due to frustrated domain growth. Magnetic imaging reveals significant changes in the domain topography when the sample is exposed to magnetic fields of different strengths. The isolated residual domains, which act as nucleation sites for the subsequent field sweep, contribute to random dipole fields that hinder domain growth, effectively setting boundaries for interconnected stripe domains. The different amounts of frustrated moments lead to the observed cross-over behavior. These results may be relevant to other systems that exhibit stripe domains.

Work at UCD has been supported by CITRIS and BaCaTec. J.E.D. and K.L. acknowledge support from a National Research Council Postdoctoral Fellowship and a UCD Chancellor's Fellowship, respectively. We thank Dr. Greg Denbeaux for technical assistance with TXRM. The ALS is supported by the Director, Office of Science, Office of Basic Energy Sciences, of the U.S. Department of Energy under Contract No. DEAC02-05CH11231.

# References

\* Corresponding author. Electronic mail: kailiu@ucdavis.edu.

[1] C. Kooy and U. Enz, Philips Res. Repts. **15**, 7 (1960).

[2] P. F. Carcia, A. D. Meinhaldt, and A. Suna, Appl. Phys. Lett. **47**, 178 (1985).

[3] D. Weller, H. Brandle, G. Gorman, C. J. Lin, and H. Notarys, Appl. Phys. Lett. **61**, 2726 (1992).

[4] C. Chappert, H. Bernas, J. Ferre, V. Kottler, J. P. Jamet, Y. Chen, E. Cambril, T. Devolder, F. Rousseaux, V. Mathet, and H. Launois, Science **280**, 1919 (1998).

[5] O. Hellwig, S. Maat, J. B. Kortright, and E. E. Fullerton, Phys. Rev. B **65**, 144418 (2002).

[6] J. Sort, B. Rodmacq, S. Auffret, and B. Dieny, Appl. Phys. Lett. **83**, 1800 (2003).

[7] Y. L. Iunin, Y. P. Kabanov, V. I. Nikitenko, X. M. Cheng, D. Clarke, O. A. Tretiakov, O. Tchernyshyov, A. J. Shapiro, R. D. Shull, and C. L. Chien, Phys. Rev. Lett. **98**, 117204 (2007).

[8] M. Seul and D. Andelman, Science **267**, 476 (1995).

[9] M. Seul and R. Wolfe, Physical Review A **46**, 7534 (1992).

[10] J. E. Davies, O. Hellwig, E. E. Fullerton, G. Denbeaux, J. B. Kortright, and K. Liu, Phys. Rev. B **70**, 224434 (2004).

[11] M. S. Pierce, C. R. Buechler, L. B. Sorensen, J. J. Turner, S. D. Kevan, E. A. Jagla, J. M. Deutsch, T. Mai, O. Narayan, J. E. Davies, K. Liu, J. H. Dunn, K. M. Chesnel, J. B. Kortright, O. Hellwig, and E. E. Fullerton, Phys. Rev. Lett. **94**, 017202 (2005).

[12] T. Thomson, G. Hu, and B. D. Terris, Phys. Rev. Lett. **96**, 257204 (2006).

[13] M. T. Rahman, N. N. Shams, Y. C. Wu, C. H. Lai, and D. Suess, Appl. Phys. Lett. **91**, 132505 (2007).

**Figure Captions**

Figure 1 -   (Color online) (a) A family of first-order reversal curves (FORC's) and (b) the corresponding FORC distribution for a [Co(4Å)/Pt(7Å)]$_{20}$ film.

Figure 2 -   (Color online) Close-up view of FORC's protruding outside of the major loop for (a) $X = 20$ and (c) $X = 50$. The corresponding switching field distribution (solid line) and DC-demagnetization curve (dashed line) are shown in (b) and (d) respectively. Three representative FORC's are highlighted: (i) start of the protrusion; (ii) maximum protrusion; and (iii) reversing from negative saturation.

Figure 3 -   (Color Online) Transmission x-ray microscopy images (top) for $X = 50$ taken at (a) 90 mT and (b) 96 mT after first applying fields of (a) 350 mT and (b) 306 mT, respectively. Inscribed circles isolate the same region where notable (a) interconnected and (b) fragmented domain growth occurs. Kerr micrographs (bottom) for a $X = 2$ sample taken at an applied field of -3.5 mT after applying fields of (c) 15 mT and (d) first -15 mT and then +3.5 mT.

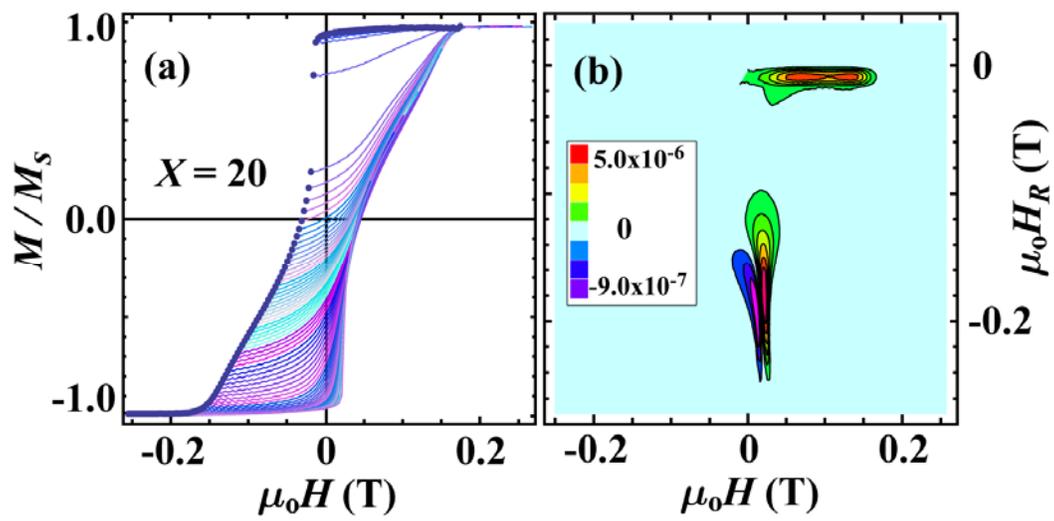

**Figure 1**

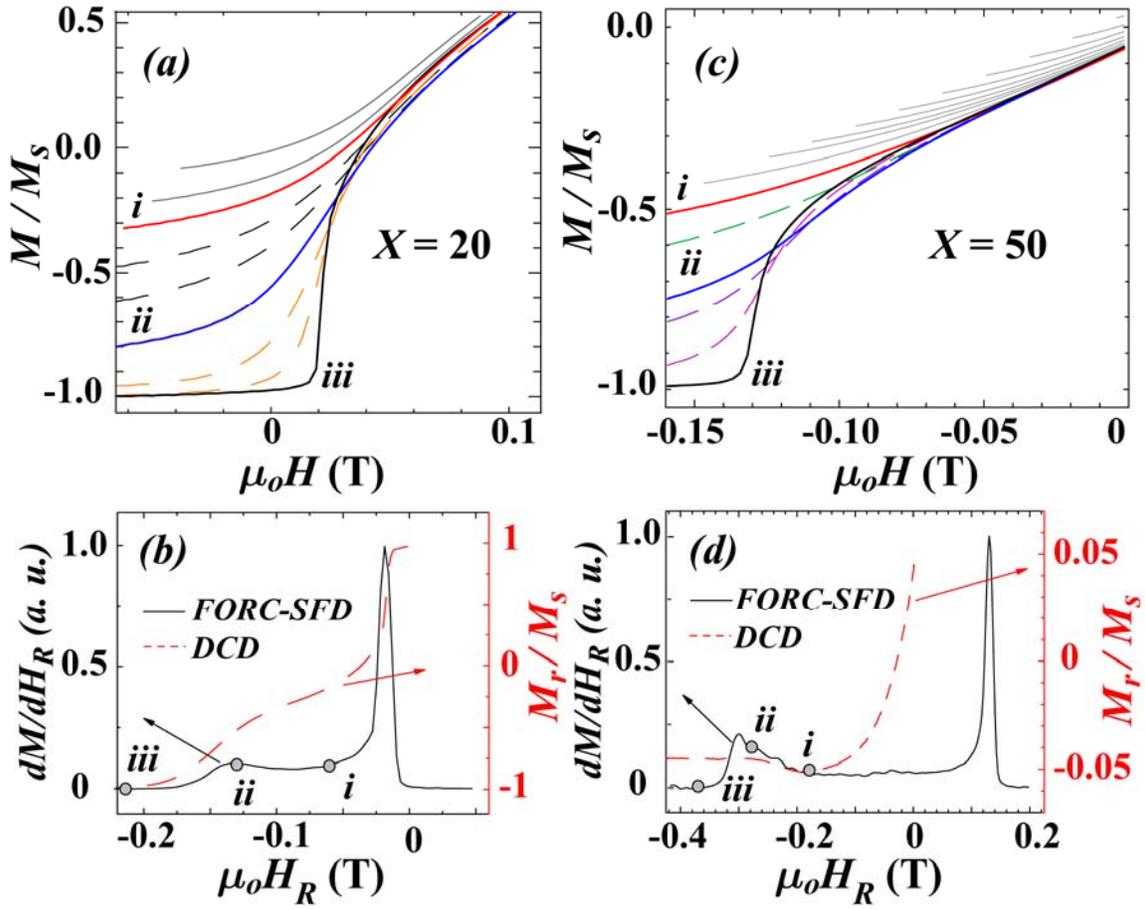

**Figure 2**

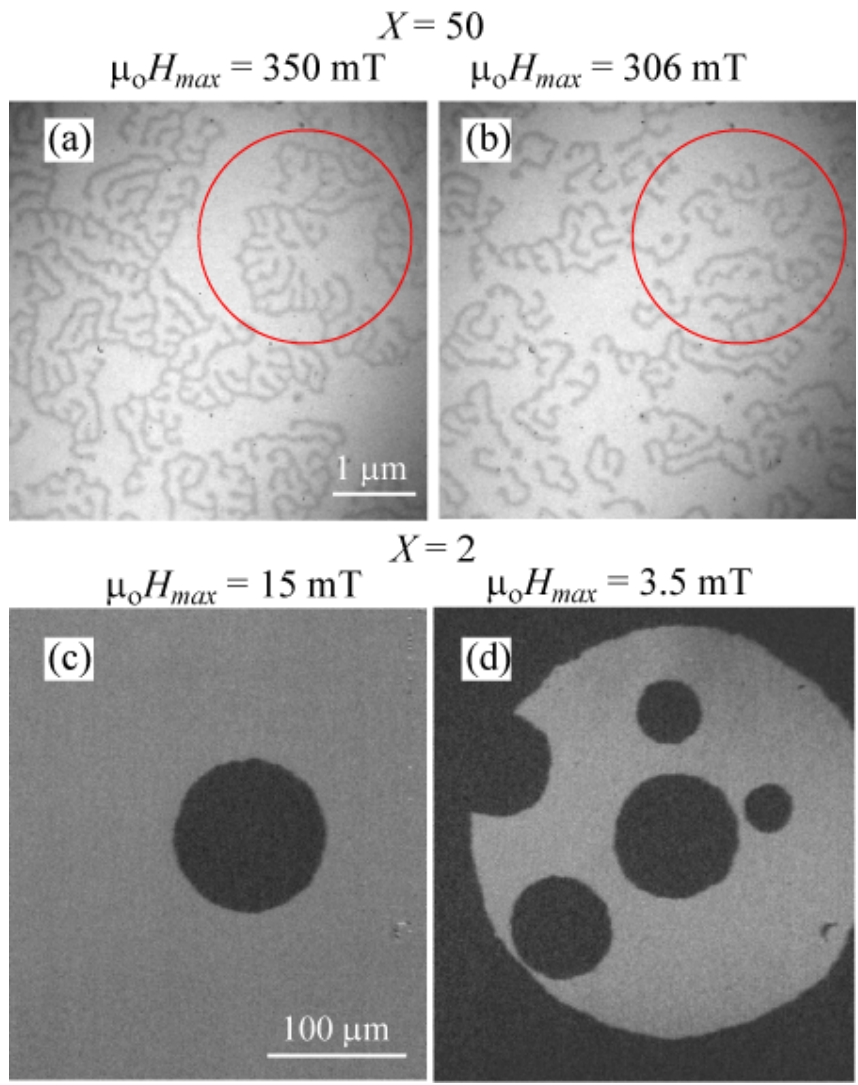

**Figure 3**